\documentclass[prl,twocolumn,showpacs,raggedbottom]{revtex4}
\usepackage{graphicx}
\begin{document}

\title{Control of Ultra-cold Inelastic Collisions by Feshbash
Resonances and Quasi-One-Dimensional Confinement}

\author{V. A. Yurovsky}
\affiliation{School of Chemistry, Tel Aviv University, 69978 Tel Aviv,
Israel}
\author{Y. B. Band}
\affiliation{Departments of Chemistry, Electro-Optics, and The
Ilse Katz Center for Nano-Science, Ben-Gurion University of the Negev,
Beer-Sheva 84105, Israel}

\date{\today}

\begin{abstract}
Cold inelastic collisions of atoms or molecules are analyzed using
very general arguments.  In free space, the deactivation rate can be
enhanced or suppressed together with the scattering length of the
corresponding elastic collision via a Feshbach resonance, and by
interference of deactivation of the closed and open channels.  In
reduced dimensional geometries, the deactivation rate decreases with
decreasing collision energy and does not increase with resonant
elastic scattering length.  This has broad implications; e.g.,
stabilization of molecules in a strongly confining two-dimensional
optical lattice, since collisional decay of the highly vibrationally
excited states due to inelastic collisions is suppressed.  The
relation of our results with those based on the Lieb-Liniger model are
addressed.
\end{abstract}


\maketitle

Feshbach resonances \cite{Feshbach_62,TTHK99} have been used to
control atomic interactions in trapped ultracold quantum gases by
tuning a magnetic field near a diatomic molecule Feshbach resonance to
convert atoms into weakly bound molecules.  For fermionic atoms the
molecules formed were remarkably long-lived \cite{Fermi_Fesh}, whereas
for bosonic atoms in a BEC \cite{Bose_Fesh}, collisional decay of the
highly excited vibrational molecular state occurs \cite{YB03} and only
a small fraction of molecules is observed in this case.

Here we show, using general scattering theory arguments, that
inelastic ultra-cold collisions in reduced dimension can be strongly
suppressed.  This generalizes the work of Ref.~\cite{Gangardt_03},
which uses the exactly solvable Lieb-Liniger many-body model for
indistinguishable bosons in one-dimension (1D) \cite{LL63}, to all
kinds of quasi-1D scattering processes occurring in atomic waveguides,
e.g., collisions of atoms and molecules.  Quasi-1D scattering occurs
in a gas in the presence of a waveguide potential that confines a 3D
gas sufficiently tightly in two directions so the radial confinement
energy $\omega_\perp$ (in units where $\hbar = 1$) is much larger than
the collision energy \cite{O98}, as in 2D optical lattices
\cite{OptLatt}, elongated atomic traps \cite{LongTrap}, and atomic
integrated optics devices \cite{AtIntOpt}.  This suppression has broad
implications, e.g., it can be used to stabilize molecules produced
from bosonic atoms in tight atomic waveguides, since
vibrational-to-translational energy-transfer collision rates are
significantly reduced relative to the 3D rates at low collision
energy.  Suppression of inelastic scattering can also occur in
collisions of other excited collision partners (e.g., in hyperfine excited
atom collisions).
Long-lived excited ultracold atoms and molecules may be useful in quantum
interferometry and quantum computation.

The theoretical framework for calculating atom-diatom scattering can
be drawn along the lines of the Arthurs and Dalgarno model
\cite{Arthurs_Dalgarno_60}.  The scattering state $| \Psi \rangle$ can
be expressed in term of a sum over basis functions,
\begin{equation}
| \Psi \rangle = \sum_j \psi_{j}({\bf r}) \,
|\chi_j \rangle ,
\end{equation}
where ${\bf r}$ is the atom-diatom relative coordinate, $\psi_{j}({\bf
r})$ is the relative wave function, and $|\chi_j\rangle$ includes
molecular and center-of-mass degrees of freedom for channel $j$.  We
shall not require details of this model since our arguments are very
general, and in fact extend beyond this particular problem (e.g.,
molecule-molecule collisions).

Low-energy inelastic exoergic collisions in the presence of a Feshbach
resonance can often be treated as multichannel scattering with
zero-range interactions described by boundary conditions for $s$-wave
radial wave functions $\varphi_{j}(r)=\frac{r}{4\pi}\int\psi_{j}({\bf
r})d\Omega_r$ (see \cite{KM02} and references therein).  In our case
of low energy inelastic scattering resulting in deactivation of the
excited state of a molecule, the boundary conditions take the form,
\begin{equation}
\left.  {d\varphi_{j}(r) \over dr} \right|_{r=0} = \sum
\limits^{}_{k=o,c,\{d\}} U_{jk}\varphi_{k}(0),
\label{phidb}
\end{equation}
for the input channel $\varphi_{o}$, the closed channel $\varphi_{c}$,
and the deactivation product $\varphi_{d}$ having a set of output
channels $\{d\}$.  This method is applicable to collisions of any type
of particles when $s$-wave scattering is allowed.  Note that broad
Feshbach molecules \cite{KGJB03} cannot be treated using the
zero-range approach of Eq.~(\ref{phidb}).  However, if one is
considering atom-molecule or molecule-molecule collisions, the
resonance does not coincide with the resonance in atom-atom
collisions, e.g., two resonances in collisions of Cs$_2$ molecules
have been observed at 12.72 and 13.15 G \cite{Chin_05}, far off the
atom-atom resonance at 19.84 G.

When the coupling of the input channel to the other channels vanishes,
Eq.~(\ref{phidb}) reduces to the Bethe-Peierls boundary condition,
\begin{equation}
\left.{d\varphi_{o}(r) \over dr}\right|_{r=0}=-{1\over a_{\text{bg}}}
\varphi_{o}(0) \mbox{ \{uncoupled\} },
\label{phibco_bg}
\end{equation}
where $a_{\text{bg}}$ is the non-resonant background elastic
scattering length.  Hence, from (\ref{phidb}), we find that
$U_{00}=-1/a_{\text{bg}}$.

Outside the interaction region, $\psi_{o}({\bf r})$ satisfies the
Schr\"odinger equation, 
\begin{equation}
-{1\over 2M}\nabla ^{2}\psi_{o}({\bf r})
+V_{\text{conf}}({\bf r}) \psi_{o}({\bf r}) = E\psi_{o}({\bf r}),
\label{psi_o}
\end{equation}
where $V_{\text{conf}}$ is the confining waveguide trapping potential,
$E$ is the collision energy, and $M$ is the reduced mass of the
colliding particles.  Moreover, the radial wave functions
$\varphi_{c}$ and $\varphi_{d}$ satisfy the Schr\"odinger equations
\begin{equation}
-{1\over 2M}{d^{2}\varphi _{c,d}\over dr^{2}}\pm
D_{c,d}\varphi_{c,d}(r) = E\varphi_{c,d}(r), 
\end{equation}
where $D_{c}$ is the asymptotic value of the closed channel potential,
$D_{d}$ is the deactivation energy for channel $d$, and we assumed
that $V_{\text{conf}}\ll D_{c,d}$.  These equations can be solved to
obtain
\begin{eqnarray}
\varphi_{c}&=&\varphi_{c}(0) \exp\left( -\sqrt{2M\left( D_{c}-E\right)
} \,r \right) \label{phi_c} , \\
\varphi_{d}(r) &=& R_{d}\exp\left( ip_{d}r\right) , \label{phi_d}
\end{eqnarray}
where $p_{d}=\sqrt{2M\left(E+D_{d}\right)}$.  The closed
channel has an attractive potential ($U_{cc}<0$) and a single bound
state with energy $E_{\text{Fesh}} = D_{c} - U^{2}_{cc} /
\left(2M\right)$.  

Substitution of Eqs.~(\ref{phi_c}) and (\ref{phi_d}) into 
(\ref{phidb}) leads to the following boundary condition:
\begin{equation}
\left.{d\varphi_{o}(r) \over dr}\right|_{r=0} = -{1\over
a_{\text{eff}}} \, \varphi_{o}(0) .
\label{phibco}
\end{equation}
The deactivation energies typically substantially exceed all
interaction energies.  Therefore only the contributions of zero
and first orders in $|U_{jk}|/p_{d}$ need be retained.  To this
accuracy we can neglect terms proportional to $U_{dd^\prime}$ in
(\ref{phidb}), so the amplitudes $R_{d}$ can be expressed as
\[
R_{d}=-i {\varphi_{o}(0) \over p_{d}} \left\lbrack U_{do}-{U_{dc}\mu
\Delta \over a_{\text{bg}}U_{oc}\left( E_{\delta}-i\Gamma _{c}\right)
}\right\rbrack ,
\]
with 
\[
\Gamma_{c}={\mu \Delta \over a_{\text{bg}}|U_{oc}|^{2}}
\sum\limits^{}_{d}{1\over p_{d}}|U_{cd}|^{2} .
\]
Here, $\Delta =a_{\text{bg}}|U_{cc}||U_{oc}|^{2}/(\mu M)$ is the
resonance strength, $\mu$ is the difference of the magnetic moments in
the closed and open channels, and
\begin{equation}
E_{\delta}={|U_{cc}|\over M}\left\lbrack \sqrt{U^{2}_{cc}+2M \left(
E_{\text{Fesh}}-E\right) }-|U_{cc}|\right\rbrack 
\end{equation}
is an effective detuning.  

The length $a_{\text{eff}}$ has an imaginary part due to coupling to
the deactivation channels and can be expressed as
\begin{equation}
a_{\text{eff}} = a_{\text{bg}}{E_{\delta}-i\Gamma_c
\over E_{\delta}+\mu \Delta-i\Gamma} ,
 \label{aresd}
\end{equation}
where
\[
\Gamma =\sum\limits^{}_{d}{1\over p_{d}}\left[ {|U_{dc}|^{2}\mu \Delta
\over a_{\text{bg}}|U_{oc}|^{2}}
+2\mu \Delta \Re\left({U_{od}U_{dc}\over U_{oc}}\right)
-a_{\text{bg}}|U_{do}|^{2}E_{\delta}\right] . \nonumber
\]
For a tightly bound closed-channel state, or when the detuning of the
collision energy from the Feshbach energy is small,
$|E_{\text{Fesh}}-E| \ll U^{2}_{cc}/M$, then $E_{\delta}\approx
E_{\text{Fesh}}-E$.  In this case, neglecting deactivation, one can
approximate $a_{\text{eff}}$ by the real effective energy-dependent
length \cite{Y05},
\begin{equation}
a_{\text{eff}}\left( E\right) \approx a_{\text{bg}}\left\lbrack 1+{\mu
 \Delta \over E-\mu \left( B-B_{0}\right) }\right\rbrack  ,
\label{aeff0}
\end{equation}
where $B-B_{0} \equiv \Delta + E_{\text{Fesh}}/\mu$ is the detuning of
the external magnetic field $B$ from its resonant value $B_{0}$.

The deactivation cross section can be expressed as
\begin{equation}
\sigma=4\pi \sum\limits^{}_{d} \varphi_{d}^*{1\over iM} {d\over dr}
\varphi_{d}
={4\pi  S \over M} |\varphi_{o}(0)|^2 ,
  \label{sigma}
\end{equation}
where $\varphi_o$ corresponds to the input channel wave function
normalized to unit incident flux density, and the factor
\begin{equation}
S=\sum\limits^{}_{d}{1\over p_d}\left |U_{do}- {U_{dc}\mu \Delta \over
a_{\text{bg}}U_{oc}\left( E_{\delta}-i\Gamma_{c}\right) } \right|^2
\label{S}
\end{equation}
describes interference of deactivation of the closed and open channel
states.

First, we consider collisions in free space ($V_{\text{conf}}=0$).
The proper solution of Eq.\ (\ref{psi_o}) has the form
\begin{equation}
 \psi_{o}({\bf r}) =\sqrt{{M\over p_{0}}}\left\lbrack \exp\left( i{\bf
 p}_{0} \cdot {\bf r}\right) -{1\over a_{\text{eff}}^{-1}+ip_{0}}
 {1\over r}\exp\left( ip_{0}r\right) \right\rbrack , \label{psi0free}
\end{equation}
where the collision momentum is $p_{0}=\sqrt{2ME}$.  For low-energy
collisions, when $a_{\text{bg}}p_{0}\ll 1$, we find
\begin{equation}
\varphi_{o}(0) = -\sqrt{{M\over p_{0}}} \, a_{\text{eff}} .
\label{psi0free0}
\end{equation}
The deactivation cross-section
\begin{equation}
\sigma_{\text{free}}={4\pi S \over p_{0}}|a_{\text{eff}}|^{2}
\label{sigma_free}
\end{equation}
diverges at low collision energies, while the deactivation rate
coefficient
\begin{equation}
K_{\text{free}} = {p_{0}\over M}\sigma_{\text{free}}\approx {4\pi S
\over M}|a_{\text{eff}}|^{2}
\label{Kfree}
\end{equation}
has a finite non-zero limit proportional to $|a_{\text{eff}}|^{2}$.
The deactivation is suppressed at $E_\delta=0$, where $a_{\text{eff}}$
is close to zero.  Under certain conditions it can also be suppressed
due to interference in the factor $S$.  The deactivation is enhanced
near resonance of $a_{\text{eff}}$ at $E_{\delta}=-\mu \Delta$.  In
this case it can reach a maximum value of
\begin{equation}
K_{\text{free,max}}\approx {4\pi S \over M }
\left({a_{\text{bg}}\mu\Delta\over\Gamma}\right)^2 .
\label{Kfree_max}
\end{equation}

Consider now collisions in a transverse harmonic waveguide.  This
problem has been analyzed in Refs.~\cite{O98,MBO04} for a
single-channel Huang pseudopotential, which is equivalent to the
boundary condition in Eq.~(\ref{phibco_bg}).  The case of a
multichannel $\delta$-function interaction has been considered in
Ref.~\cite{Y05} using a renormalization procedure.  Equations (17) and
(19) in \cite{Y05} express the proper solution of Eq.\ (\ref{psi_o})
in terms of the transverse Hamiltonian eigenfunctions $|n0\rangle$
with zero angular momentum projection on the waveguide axis, $z$,
\begin{eqnarray}
\psi_{o}({\bf r}) && = a_\perp\sqrt{{\pi M\over p_{0}}}\biggl\lbrack
\exp\left( ip_{0}z\right) |00\rangle - {1\over 2}Ma_{\perp }
\nonumber \\
&& \times T_{\text{conf}}\left( p_{0}\right)
\sum\limits^{\infty }_{n=0}{\exp\left( ip_{n}|z|\right) |n0\rangle \over
\sqrt{n-\left( p_{0}a_{\perp }/2\right)^{2}}}\biggr\rbrack .
\label{psi0conf}
\end{eqnarray}
Here $a_{\perp }=\left( M\omega_{\perp }\right) ^{-1/2}$ is the
transverse harmonic oscillator length, $\omega_{\perp }$ is the
waveguide transverse frequency, $p_{n}=\sqrt{2M\left\lbrack E - \left(
2n+1 \right) \omega_{\perp }\right\rbrack }$,
\begin{equation}
T_{\text{conf}}\left( p_{0}\right) ={2\over Ma_{\perp }}\left\lbrack
{a_{\perp }\over a_{\text{eff}}}+\zeta \left( {1\over 2},-\left(
{a_{\perp }p_{0}\over 2}\right) ^{2}\right) \right\rbrack ^{-1}
\label{Tconf}
\end{equation}
is the transition matrix, and $\zeta(\nu ,\alpha)$ is the Hurwitz zeta
function \cite{MBO04,BE53}.  The wave function (\ref{psi0conf}) is
normalized so the average incident flux density per waveguide area
$\pi a_\perp^2$ is unity.  The sum in Eq.~(\ref{psi0conf}) diverges as
$r\rightarrow 0$.  The divergent part can be evaluated as $a_{\perp
}/r$ \cite{MBO04}.  This leads to $\varphi_{o}(0) = -{1\over
2}Ma_{\perp }^2\sqrt{M/p_{0}} T_{\text{conf}} \left(
p_{0}\right)$, and to the deactivation rate coefficient
\begin{equation}
K_{\text{conf}} =\pi M a_{\perp }^4 |T_{\text{conf}}|^2 S.
\label{Kconf}
\end{equation}

For weak confinement, when $a_{\perp }p_{0}\gg 1$, 
approximation (49) in Ref.~\cite{Y05} leads again to Eq.\ (\ref{psi0free})
for the wave function and to Eq.\ (\ref{Kfree}) for the deactivation rate.
For strong confinement, when $a_{\perp }p_{0}\ll 1$, approximation
(41) in Ref.~\cite{Y05} leads to
\begin{equation}
T_{\text{conf}}( p_0) \approx -i{p_0\over M} 
\left( 1+{i\over 2}Ca_{\perp}p_{0}
-i{a^{2}_{\perp }p_{0}\over 2 a_{\text{eff}}}\right) ^{-1} ,
\end{equation}
where $C \approx 1.4603$.  At low collision energies, or at large
$a_{\text{eff}}$, where $ p_{0} \ll |a_{\text{eff}}|/a^{2}_{\perp } $,
the wave function at the origin, $\varphi_{o}(0)\approx{i\over 2}
a^{2}_{\perp }\sqrt{M p_0}$, is much less than the corresponding value
of $\varphi_{o}(0)$ in free space given in Eq.~(\ref{psi0free0}).
Thus confinement prevents the particles from occupying the same
position.  A similar effect is responsible for fermionization of 1D
bosons with strong interactions \cite{O98}.  Under these conditions
the deactivation rate,
\begin{equation}
K_{\text{conf}}\approx {a^{4}_{\perp}
p^{2}_{0}\over 4|a_{\text{eff}}|^{2}}K_{\text{free}} ,
\end{equation}
can be substantially suppressed by confinement.

\begin{figure}
\includegraphics[scale=0.4]{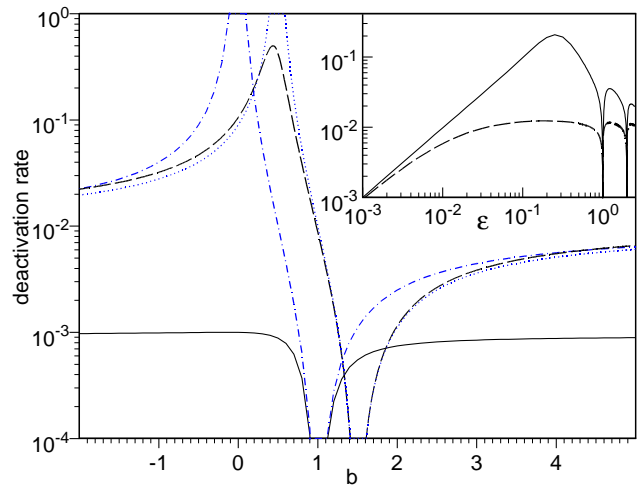}
\caption{(color online) Scaled deactivation rate coefficient $\frac{K
M}{4\pi S a^2_\perp}$ as a function of scaled magnetic field detuning
$b=\mu(B-B_0)/(2\omega_\perp)-{1\over 2}$ and collision energy
$\varepsilon=E/(2\omega_\perp)-{1\over 2}$, calculated for
$a_{\text{bg}}=0.1 a_\perp$ and $\mu\Delta=2\omega_\perp$.  In the
confined geometry, the solid and dashed curves correspond to
$\varepsilon=10^{-3}$ and $\varepsilon=0.5$ respectively, whereas the
free space results are given by the dot-dashed and dotted curves
respectively.  The inset shows deactivation rate versus $\varepsilon$
in confined geometry for $b=0$ (solid curve) and $b=100$ (dashed
curve).}
\label{Fig_K}
\end{figure}

This conclusion is graphically demonstrated in Fig.~\ref{Fig_K} under
conditions when $a_{\text{eff}}$ is expressed by Eq.~(\ref{aeff0}).
It shows resonances in the deactivation rate at $E_\delta=-\mu\Delta$
for collision energies comparable to $\omega_\perp$ and in free space,
as well as deactivation suppression near $E_\delta=0$.  At low
collision energies, when
\begin{equation}
p_{0}\ll |a_{\text{bg}}|/a^{2}_{\perp }, \label{suppr_cond}
\end{equation}
deactivation under confinement does not have resonances and can be
strongly suppressed even compared to the non-resonant process in free
space.  Suppression appears also at $E=(2k+1)\omega_\perp$, where
excitations of transverse waveguide modes become open, leading to
jumps in the elastic scattering amplitude \cite{Y05,GB04}.

The above results are obtained for a system composed of two arbitrary
types of particles interacting via $s$-wave scattering.  A suppression
of inelastic collision has been predicted in Ref.~\cite{Gangardt_03}
for a many-body system of 1D indistinguishable bosons using the
Lieb-Liniger model \cite{LL63}.  However, as we shall see below, the
suppression is mostly a two-body interaction effect even in this
model.

Consider first the two-body scattering process with particle momenta
$p_{1}$ and $p_{2}$.  The Lieb-Liniger wave function with
unit norm in interval $\lbrack 0,L\rbrack $ ($L\rightarrow \infty $)
is,
\begin{eqnarray}
\Psi ^{\left( 2\right) }_{p_1p_2}\left( z_{1},z_{2}\right) ={1\over
\sqrt{2}L}\biggl\lbrack \exp\left( ip_{1}z_{1}+ip_{2}z_{2}\right)
\nonumber \\
+ {p_{1}-p_{2}-imU_{a}\over p_{1}-p_{2}+imU_{a}}\exp\left(
ip_{1}z_{2}+ip_{2}z_{1}\right) \biggr\rbrack ,
\nonumber
\end{eqnarray}
for $0\le z_{1}\le z_{2} < L$ and $\Psi ^{\left( 2\right)
}_{p_1p_2}\left( z_{1},z_{2}\right) =\Psi ^{\left( 2\right)
}_{p_1p_2}\left( z_{2},z_{1}\right) $ for $0\le z_{2}\le z_{1}<L$.
Here $m$ is the particle mass, and $U_{a}\approx 4a_{\text{bg}}\left[
ma^{2}_{\perp }(1-C a_{\text{bg}}/ a_{\perp })\right] ^{-1}$ is the
interaction strength \cite{O98}.  The two-body correlation function
with the particles at the same position,
\begin{equation}
 g^{\left( 2\right) }_{2}\left( p_{1},p_{2}\right) = \left| \Psi
 ^{\left( 2\right) }_{p_1p_2}\left( 0,0\right) \right|^2 = {2\over
 L^{2}} {\left( p_{1}-p_{2}\right)^{2}\over \left( p_{1}-p_{2}\right)
 ^{2} +m^{2}U^{2}_{a}}
\label{g2_2}
\end{equation}
is the probability to find two particles at the same place.  Equation
(\ref{g2_2}) already describes qualitatively the behavior of $g_2$
when the ratio of the interaction to collision energies is large, as
obtained in \cite{Gangardt_03}, $g_2\sim(p_1-p_2)^2/U_a^2$.

In the $N$-body case, the two-body correlation function $g^{(N)}_2$ 
can be estimated as a sum of $g^{\left( 2\right)
}_{2}$ over all pairs of the colliding particles with the quasimomenta
$p_{j}$ and $p_{j^\prime }$,
\begin{eqnarray}
g^{\left( N\right) }_{2} &\approx& \sum\limits^{}_{j<j^\prime }
g^{\left(2\right)}_{2} (p_{j},p_{j^\prime }) \nonumber \\
&\approx& {L^{2}\over 2}\int dp_{1}dp_{2}f(p_{1})
f(p_{2}) g^{\left(2\right)}_{2}( p_{1},p_{2})
, \label{g2N}
\end{eqnarray}
where the values of the quasimomenta $p_{j}$ are determined by
boundary conditions and the summation is replaced by integration with
the quasimomentum distribution functions $f(p)$ \cite{LL63}.  The
system properties are determined by the dimensionless parameter
$\gamma = mU_{a}/\rho$, where $\rho=N/L$ is the linear particle
density.  Approximate analytical expressions for $f(p)$ in the ground
state have been obtained in Ref.~\cite{LL63} for the mean-field
regime, $\gamma \ll 1$, and for the Tonks-Girargeau regime, $\gamma
\gg 1$.  In the mean-field regime, substitution of $f(p) \approx
\pi^{-1} \gamma^{-1/2} \sqrt{1-p^{2} /(4\rho^{2}\gamma )}$ into
Eq.~(\ref{g2N}) leads to
\begin{equation}
g^{\left( N \right)}_{2}\approx \rho^{2} ,
\end{equation}
in full agreement with the results of Ref.~\cite{Gangardt_03}.  In the
Tonks-Girardeau regime, $f(p) \approx 1 /(2\pi) $ for $|p|<\pi\rho $
and $f(p) = 0$ otherwise, and (\ref{g2N}) leads to
\begin{equation}
g^{\left( N\right) }_{2}\approx {2\pi ^{2}\rho^{2}\over 3\gamma^{2}} .
\label{g2NTG}
\end{equation}
This value is half the exact value determined in
Ref.~\cite{Gangardt_03}.  The difference can be related to higher
order correlations, which are neglected here (the two-body picture can
only include second order correlations).  However, (\ref{g2NTG})
describes the correct behavior of $g^{\left( N\right) }_{2}$ as
$\gamma \rightarrow \infty$, leading to suppression of all kinds of
collision phenomena under tight confinement when $mU_{a}/\rho >> 1$
[this condition has the same meaning as Eq.\ (\ref{suppr_cond})].

In summary, inelastic collision rates in free space demonstrate
resonances and dips, being proportional to $|a_{\text{eff}}|^{2}$, and
are capped by (\ref{Kfree_max}).  Interference of deactivation of the
open and closed channels can also suppress the rate.  In quasi-1D
scattering at low collision energies [see Eq.~(\ref{suppr_cond})],
inelastic collisions do not have resonances and are suppressed.  This
result applies to the collision of any type of atoms or molecules
interacting via $s$-waves and is not based upon the 1D Bose gas
Lieb-Liniger model.

\begin{acknowledgments}
This work was supported in part by grants from the U.S.-Israel
Binational Science Foundation (grant No.~2002147), the Israel Science
Foundation for a Center of Excellence (grant No.~8006/03), and the
German Federal Ministry of Education and Research (BMBF) through the
DIP project.  Useful conversations with Paul Julienne and Brett Esry
are gratefully acknowledged.
\end{acknowledgments}

\end{document}